\documentclass[conference, 9pt]{IEEEtran}
\usepackage{cite}
\usepackage{amsmath,amssymb,amsfonts}
\usepackage{algorithmic}
\usepackage{breqn}
\usepackage{bm}
\usepackage{caption}
\usepackage{comment}
\usepackage{graphicx}
\usepackage{makecell}
\usepackage{multirow}
\usepackage{soul}
\usepackage{subcaption}
\usepackage{textcomp}
\usepackage{url}
\usepackage{xcolor}
\usepackage{hyperref}
\def\BibTeX{{\rm B\kern-.05em{\sc i\kern-.025em b}\kern-.08em
    T\kern-.1667em\lower.7ex\hbox{E}\kern-.125emX}}

\begin{document}

\title{{\system{}}: Automated Discovery of New Classes in Audio Datasets}
\begin{comment}
\name{\centering\begin{tabular}{ccc}
      Ryuhaerang Choi$^{\star}$\sthanks{Most of this work was done during the author's internship at Nokia Bell Labs.}  \hspace*{.5cm} & Soumyajit Chatterjee$^{\dagger}$ \hspace*{.5cm} &  Dimitris Spathis$^{\dagger}$  \\
      Sung-Ju Lee$^{\star}$ \hspace*{.5cm} & Fahim Kawsar$^{\dagger \ddagger}$ \hspace*{.5cm} &  Mohammad Malekzadeh$^{\dagger}$
      \end{tabular}}
\address{$^{\dagger}$Nokia Bell Labs Cambridge, UK \hspace*{.5cm} $^{\star}$KAIST, South Korea 
 \hspace*{.5cm} $^{\ddagger}$University of Glasgow, UK }
\end{comment}

% \sthanks{Most of this work was done during the author's internship at Nokia Bell Labs.}

\author{
	\IEEEauthorblockN{Ryuhaerang Choi$^*$\IEEEauthorrefmark{2}, Soumyajit Chatterjee\IEEEauthorrefmark{3}, Dimitris Spathis\IEEEauthorrefmark{3}, Sung-Ju Lee\IEEEauthorrefmark{2}, 
 \\Fahim Kawsar\IEEEauthorrefmark{3}\IEEEauthorrefmark{4}, Mohammad Malekzadeh\IEEEauthorrefmark{3}}
	\IEEEauthorblockA{\IEEEauthorrefmark{3}Nokia Bell Labs, Cambridge, UK \IEEEauthorrefmark{2}KAIST, South Korea  
 \IEEEauthorrefmark{4}University of Glasgow, UK}
    
	Email: \{ryuhaerang.choi, profsj\}@kaist.ac.kr,\\
	
	\{soumyajit.chatterjee, dimitrios.spathis, fahim.kawsar, mohammad.malekzadeh\}{@nokia-bell-labs.com}}

\newcommand{\red}[1]{\textcolor{red}{#1}}
\definecolor{lavender}{rgb}{0.86, 0.81, 1}
\newcommand{\hr}[1]{\sethlcolor{lavender}\hl{[HR: #1]}}
\newcommand{\ts}[1]{\sethlcolor{green}\hl{[Sumo: #1]}}
\newcommand{\ds}[1]{\sethlcolor{yellow}\hl{[Dimitris: #1]}}
\definecolor{champagne}{rgb}{0.97, 0.91, 0.81}
\newcommand{\mo}[1]{\textcolor{teal}{[Mo: #1]}}
\newcommand{\newadd}[1]{\textcolor{black}{#1}}
\newcommand{\hredit}[1]{\textcolor{black}{#1}}

\newcommand{\system}{{\textsl{SoundCollage}}}

\maketitle

\begingroup\renewcommand\thefootnote{}
\footnotetext{\IEEEauthorrefmark{1}Work is mainly done during the author’s internship at Nokia Bell Labs.}
\endgroup

\begin{abstract}  
% 100-150 word counts
Developing new machine learning applications often requires the collection of new datasets. However, existing datasets may already contain relevant information to train models for new purposes. We propose \system{}: a framework to discover \emph{new classes} within audio datasets by incorporating (1)~an {\em audio pre-processing pipeline} to decompose different sounds in audio samples, and (2)~an automated {\em model-based annotation mechanism} to identify the discovered classes. Furthermore, we introduce the \emph{clarity} measure to assess the coherence of the discovered classes for better training new downstream applications. Our evaluations show that the accuracy of downstream audio classifiers within discovered class samples and a held-out dataset improves over the baseline by up to $34.7\%$ and $4.5\%$, respectively. These results highlight the potential of \system{} in making datasets reusable by labeling with newly discovered classes. To encourage further research in this area, we open-source our code at \textcolor{blue}{\href{https://github.com/nokia-bell-labs/audio-class-discovery}{github.com/nokia-bell-labs/audio-class-discovery}}.
\end{abstract}

\section{Introduction}
\label{intro}
Training or fine-tuning a machine learning~(ML) model for emerging audio processing applications often requires high-quality labeled data. This necessity is typically fulfilled by collecting a new audio dataset tailored to the intended application, ensuring that the semantics of the annotated labels align with the application's needs. Notably, audio signals often capture rich contextual and environmental information. For instance, datasets collected for Coronavirus disease detection through coughing sounds may also contain other respiratory sounds considered irrelevant for the main  application~\cite{xia2021covid}. Such additional audio signatures can be \textbf{reused} to train models for other healthcare applications like detecting chronic obstructive pulmonary disease or heart diseases~\cite{baur2024hear}. In other words, when processed effectively, audio datasets can often accommodate the requirements of multiple applications, reducing the costs for a separate data collection.

%%%%%%%%%%%
This presents an opportunity to explore beyond the annotated label space and discover \emph{new classes} within existing datasets. To identify new classes at runtime, prior works have mostly adapted zero- (or few-) shot learning~\cite{han2019learning,troisemaine2023novel}, which identify unknown classes using the existing definitions available from the known classes. Other methods with supervised clustering rules also work similarly by confining the search for new classes based on existing classes~\cite{shu2018unseen,chi2022meta} or known musical components~\cite{gha2023unsupervised, wang2024mir}. Naturally, all these approaches limit the search for new classes to semantic attributes of the known classes, such as using predefined audio signatures or signal characteristics derived from domain knowledge, which can be challenging to obtain in many real-world scenarios. Finally, other supervised methods~\cite{wu2020automated} require costly supervision by engaging users during runtime to specify the class to which the current audio signal belongs, which may be an unrealistic assumption.

%%%%%%%%%%
\newadd{%The most promising approach is \textbf{unsupervised task-discovery}, first proposed in~\cite{atanov2022task}, where new class boundaries are discovered within data that may not have been labeled beforehand. 
In contrast, unsupervised task discovery, first introduced in~\cite{atanov2022task}, presents a promising direction by discovering new class boundaries within data that may not have been labeled beforehand. This approach offers a flexible solution, utilizing inherent inductive biases and statistical patterns present within the data by using ML models as a proxy. However, the current unsupervised task-discovery approaches do not assign meaningful labels to the discovered classes. The major issue is that all existing works~\cite{chi2022meta, Hsu18_L2C,atanov2022task} are tailored for image data only and do not accommodate the nuances of audio data, which require fundamentally different inductive biases given the natural differences in the data properties compared to images~\cite{zhang2019deep}. Our systematic investigation shows~(see \tablename~\ref{tbl:diversity}) that the existing baseline has limited capability in discovering unique auditorily semantic classes from audio datasets.}%fails to discover unique classes from the audio datasets.}

\begin{figure}[]
    \centering
    \includegraphics[width=\columnwidth]{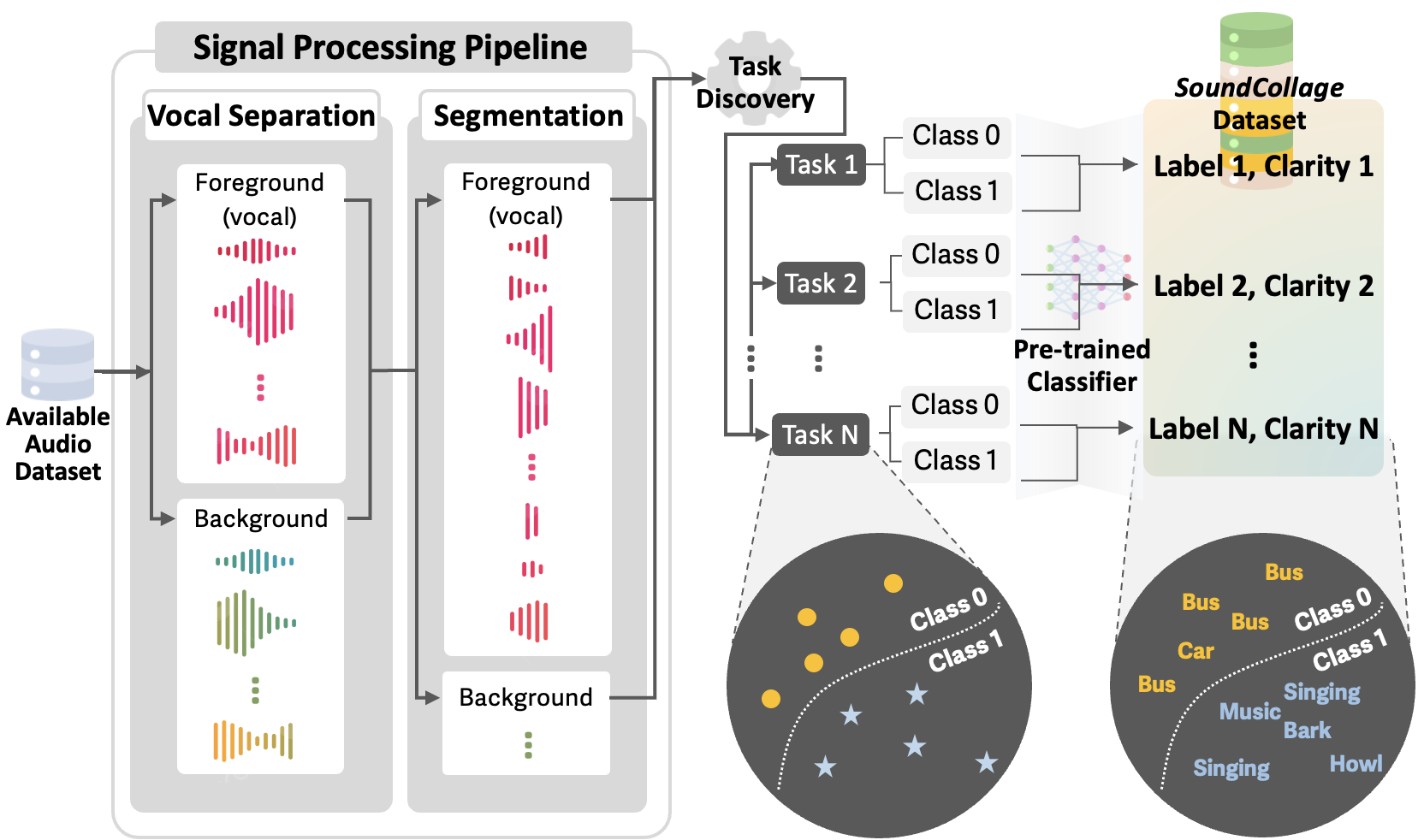} 
    \caption{\textbf{Conceptual illustration of \system{}}.}
    \label{fig:taskovery}
\end{figure}

%%%%%%%%%%%
We present \system{}, a framework to discover classes within existing audio datasets without direct human supervision. \newadd{As illustrated in~\figurename~\ref{fig:taskovery}, we first employ a set of signal processing steps designed to decompose the different components of audio signals.} These steps reduce the inherent complexity of audio signals in which various patterns, such as human voice elements and background noises, intertwine. The decomposed components are then fed into a subsequent unsupervised task discovery module to identify new classes. \system{} then automatically annotates these identified classes with human-readable labels, for which we utilize pre-trained audio-event classification models, such as YAMNet~\cite{yamnet}. \newadd{The final output of our proposed framework is a ``\emph{collage of sounds}'' labeled with different newly discovered classes. To summarize our contributions, %compared to existing unsupervised class discovery approaches, 
we (1)~propose a set of systematically selected audio signal processing approaches to decompose the complex audio data, (2)~incorporate pre-trained models to label the newly discovered classes, (3)~introduce a new measure, \emph{clarity}, to evaluate the semantic coherence of discovered classes, and finally, (4)~train models with the newly discovered classes, thus, \emph{reusing} the same dataset for new applications, albeit originally collected for different purposes.}

We evaluate \system{} on AudioSet~\cite{AudioSet}, a multi-label benchmark dataset for audio-event detection that includes a diverse set of classes. Using our proposed clarity measure, we conduct a systematic evaluation to assess the quality of the identified class boundaries. \system{} outperforms the baseline~\cite{atanov2022task}, achieving an average 
clarity improvement of $1.3\%$.
% \hredit{clarity improvement of 0.013 ($\pm$ 0.016) (on a scale from 0 to 1).}
%((0.306-0.302)+(0.286-0.302)+(0.350-0.336)+(0.367-0.336)+(0.372-0.342)+(0.355-0.342))/6 = 0.013  
Importantly, the gain in clarity of the class boundaries is reflected in the performance of downstream classifiers. Our evaluations on a held-out test set obtained from FSD50K~\cite{fonseca2022FSD50K}, an additional benchmark audio dataset, demonstrate that the downstream classifiers trained using the data discovered by \system{} achieve an accuracy improvement up to $4.5\%$. 

\section{Methodology}
\label{method}
{\bf Problem Statement.}
We assume a dataset $\mathcal{D}=<\mathcal{A},\mathcal{L}>$ is given where $\mathcal{A}$ is the set of audio samples and $\mathcal{L}$ \newadd{is the \emph{optional} label information available from initial human annotations.} We aim to discover new classes $c_1, c_2, \dots$ such that $c_i \notin \mathcal{L}$, and automatically assign a semantic label to each $c_i$ without human supervision.
\subsection{Signal Pre-processing Pipeline}
The primary feature of audio data is the presence of multiple sound sources, making the underlying audio signatures a convoluted signal of all the individual audio components. For example, audio data with domestic sounds can contain sounds caused by domestic activities, pets, adult human speech, child cry, child speech, etc.~\cite{9746096}. Inspired by this feature, \system{} separates key acoustic signal components from an audio signal to efficiently discover new classes from the data. We achieve this by designing signal pre-processing steps which aim to generate augmented samples with fewer concurrent sound sources. Without pre-processing, the complexity of original audio could complicate identifying common acoustic patterns across multiple samples for discovering new classes. Thus, the objective of this step is to decompose a complex audio signal into simpler acoustic components.
% Without \system{}'s pre-processing, the convoluted nature of the original data may make the subsequent step of identifying common acoustic patterns across multiple samples to discover a new class extremely challenging. Thus, a vital objective of this step is to break down complex audio samples into simpler acoustic components.

We first employ \emph{vocal separation} to separate overlapping vocal and background sounds within samples using the REPET-SIM method~\cite{rafii2012music}. Unlike the original binary mask approach, we use Wiener filtering to generate a soft mask and adjust the FFT window's overlap accordingly. \newadd{Notably, Wiener filtering is widely utilized in human voice processing due to its optimal performance in minimizing the mean squared error (MSE) between the estimated and desired signals. Its ability to adapt to the statistical characteristics of both signal and noise makes it exceptionally suited for dynamically varying noise environments typical in voice communications, allowing us to separate the voice and background signatures into separate components~\cite{7177993}.} Specifically, each sample is divided into two components:~one with enhanced vocal components (Comp\#1), and the other containing background audio signatures (Comp\#2).

However, each of these components can still contain significant variations. For example, the vocal component can include alternating sequences of adult and child voices. \newadd{Similar observations are also present from the background audio where multiple sound sources can be present.} Building on this insight, we further \emph{segment} the data by applying a change-point-detection algorithm~\cite{arlot2019kernel} to divide continuous audio streams, ensuring each segment captures homogeneous acoustic patterns by avoiding abrupt time-domain changes. %\newadd{The output of this step is a \textbf{``collage of sounds''} that potentially contains audio samples with segregated sound signatures.} 
% \hr{I used the term `collage' to refer to the final stage, actually. The definition of `collage' is `\textit{a piece of art made by sticking various different materials such as photographs and pieces of paper or fabric onto a backing.}' In ouf case, the new classes we obtain at the end are like meaningful works of art (as in the definition of `\textit{a piece of art}') formed by taking pieces from the data samples (as in `\textit{pieces of paper or fabric}' from the definition) and putting them together. Given this definition, I'm having trouble understanding why it's referred to as a `collage of sounds,' here. What are you thoughts on this?}
This method facilitates further discovery of classes by exploiting consistent patterns within each segment. Remarkably, this segmentation also increases the total number of samples for the subsequent class discovery.

\subsection{Task Discovery}
\newadd{Task discovery is based on the principle that when labels in a dataset are meaningful, DNNs with the same architecture but different initializations, trained on this dataset, show notable similarities in their output space~\cite{hacohen2020let,10.1007/978-3-031-20074-8_23}. 
To formulate this, let \(\tau\) denote a {\em task} defined by binary labels assigned to a set of audio samples \( X \), i.e.,~\( \tau: X \rightarrow \{0, 1\} \). Let \( \mathcal{D}(X, \tau) = \{(x,\tau(x))|x\in X\} \) denote a labeled dataset for task \(\tau\). Let \(\mathcal{P}(\cdot)\) denote a learning algorithm, e.g., stochastic gradient descent using binary cross-entropy loss, and \( w\sim\mathcal{P}(\mathcal{D}(X,\tau)) \) denote the outcome of training a randomly initialized model for task \(\tau\). To assess the generalization of \( w \), we split \( \mathcal{D}(X, \tau) \) into a training set \( \mathcal{D}(X_{tr}, \tau) \) and a test set \( \mathcal{D}(X_{te}, \tau) \). After training two DNNs with different initializations, \( w_1 \) and \( w_2\), \textit{agreement score}~(AS)~\cite{pnas1903070116, atanov2022task} 
% on the output space 
is defined as:
\begin{equation}
    \begin{split}
        AS(\tau; X_{tr}, X_{te}) =
    \mathbb{E}_{w_1,w_2\sim \mathcal{P}(\mathcal{D}(X_{tr},\tau))}\mathbb{E}_{x\sim X_{te}} \\
        [(f(x; w_1) = f(x; w_2)].
    \end{split}
    \label{eqn:as}
\end{equation}
AS is used to discover new class boundaries within a dataset of images~\cite{atanov2022task}. First, a task \(\tau\) is generated by randomly assigning a label~(either 0 or 1) to each sample in the dataset. Next, to estimate the AS value in Eqn.~\ref{eqn:as}, multiple iterations are performed, where in each iteration, two randomly initialized DNNs are trained on \(\tau\). The task discovery algorithm then finds a task \( \tau \) which maximizes \( AS(\tau) \).  Finally, a task \(\tau\) is identified as a new class boundary if the AS is higher than a threshold, as a high AS indicates meaningful class boundaries that closely resemble human-labeled tasks.}

\subsection{Automated Labeling of Discovered Classes}
After discovering a set of tasks \(\tau_1, \dots, \tau_K\) with high AS, we need to translate each task into a meaningful class \(c_1, \dots, c_K\). A simplistic approach is to rely on costly human annotators to label the discovered tasks~\cite{wu2020automated, acconotate}, which is especially unrealistic when dealing with audio datasets collected from real-world environments that contain numerous hidden classes. To this end, we utilize an existing \textit{pre-trained audio-event classification} model to assign meaningful classes to the discovered tasks. For each discovered task \( \tau_i \), we consider $N_i$ samples from the binary classes present within the task. These samples are then passed to the pre-trained  model to assign the new class
\( c_i \) to the task \( \tau_i \). The output of this step is a set of \textbf{``sound collages''} which is composed of pieces from different audio sources grouped into \textbf{newly discovered, meaningful classes}.

\subsection{Clarity Measure}\label{sec:clarity}
A significant drawback of existing unsupervised task discovery work~\cite{atanov2022task} is the lack of a measure to assess how clearly one class differentiates from another from a {\em semantic} perspective, as if evaluated by humans. Current baseline solely relies on the model's perspective, assessed by AS, which only measures how well a trained model generalizes on a given task. To fill this gap, we introduce the \textbf{clarity} measure, which enables principled evaluation of class boundaries from a semantic standpoint. The details follow. 

Given a task $\tau_i$ with $N_i$ samples, let $Y_j$ represent a specific label from the set of semantic labels obtained for all samples, using YAMNet~\cite{yamnet} (e.g., $Y_j$: singing). The total number of samples in class~0 and class~1 are denoted as $N^0_i$ and $N^1_i$, respectively. The number of samples in class~0 labeled with $Y_j$ is denoted as $N^0_{ij}$~(similarly in class 1, it is denoted as $N^1_{ij}$). The clarity $C$ of a class boundary is defined as
\begin{equation}
    % \scriptsize
    C = \max\bigg(\frac{|N^0_{ij} - N^1_{ij}| - min\{N^0_{ij}, N^1_{ij}\}}{max\{N^0_i, N^1_i\}}, 0\bigg).
    \label{eqn:clarity}
\end{equation}

Eqn.~\eqref{eqn:clarity} measures the quality of the class boundary discovered in a task for a given semantic label. To clarify, consider two tasks $\tau_1$ and $\tau_2$ with their corresponding class boundaries for a YAMNet label gram. Suppose $\tau_1$ for label $Y_0$ has fifteen samples in class 1 and five in class 0, while  $\tau_2$ for label $Y_0$ has ten samples in class 1 and none in class 0. Given the label $Y_0$, Eqn.~\eqref{eqn:clarity} penalizes $\tau_1$ for its less distinct boundary while highlighting $\tau_2$ for higher clarity.
% and highlights $\tau_2$ for having a more precise class boundary.

% A fundamental challenge following the discovery of class boundaries is to discern the semantics underlying these boundaries.
% Conventional solutions for identifying these classes often rely on human annotators, asking them to label the unidentified classes discovered by the system~\cite{wu2020automated}. However, engaging human annotators is often impractical and costly~\cite{acconotate}, especially when dealing with audio datasets collected from real-world environments~\cite{ubicoustics}, which often contain numerous hidden classes.
% Given these considerations, we propose an end-to-end solution that provides annotations for the discovered classes without the need for human intervention. We accomplish this by incorporating a pre-trained audio event classification model to assign semantic meanings, or labels, to the discovered classes. %We accomplish this by incorporating a pre-trained audio event classification model. To get the labels for the discovered classes, t
% \system{} looks into the discovered tasks and considers $N_i$ samples from the binary classes present within a task. These samples are then passed to the pre-trained audio event classification model to obtain the labels. These labels can be used either as pseudo-labels directly or as interpretative references for the discovered classes.
\section{Evaluation}
\label{eval}

\subsection{Datasets}
\label{sec:datasets}
%\subsection{For Discovery of New Classes}

\begin{table}[h]
\caption{\textbf{Details of the processed AudioSet.} Our pre-processing creates two components from each sample: Comp\#1 with enhanced vocal signatures, and Comp\#2 with background audio signatures.} 
\label{tab:datasets}
\resizebox{\columnwidth}{!}{
\scriptsize
\begin{tabular}{crcrr}
\Xhline{2\arrayrulewidth}
\multirow{2}{*}{\textbf{Class}} & \multicolumn{1}{c}{\multirow{2}{*}{\textbf{\begin{tabular}[c]{@{}c@{}}\# of original\\ samples\end{tabular}}}} & \multicolumn{3}{c}{\textbf{\# of samples after signal pre-processing}} \\ \cline{3-5} 
 & \multicolumn{1}{c}{} & \multicolumn{1}{l}{} & \multicolumn{1}{c}{\textbf{Total}} & \multicolumn{1}{r}{\textbf{Selected}} \\ \hline\hline
\multirow{2}{*}{\textbf{Speech}} & \multirow{2}{*}{797} & Comp\#1 & 2,288 & 2,288 \\
 &  & Comp\#2 & 2,164 & 2,164 \\ \hline
\multirow{2}{*}{\textbf{\begin{tabular}[c]{@{}c@{}}Domestic sounds,\\ home sounds\end{tabular}}} & \multirow{2}{*}{8,237} & Comp\#1 & 21,895 & 10,000 \\
 &  & Comp\#2 & 21,323 & 10,000 \\ \hline
\multirow{2}{*}{\textbf{\begin{tabular}[c]{@{}c@{}}Outside rural\\ or natural\end{tabular}}} & \multirow{2}{*}{12,463} & Comp\#1 & 34,278 & 10,000 \\
 &  & Comp\#2 & 33,020 & 10,000 \\ \Xhline{2\arrayrulewidth}
\end{tabular}
}
\end{table}

\newadd{We run \system{} on AudioSet~\cite{45857}, a multi-label benchmark dataset with} 632~audio event classes organized in a hierarchical ontology.  We focus on classes higher up in the hierarchy as they are more likely to contain a mixture of various sounds. Notably, the annotations in AudioSet vary in quality. This in turn often results in a drop in overall quality of the dataset with audio data incorrectly labeled for the presence of certain audio signatures, such as sounds from animals or vehicles~\cite{damiano2024can}. Therefore, \newadd{for a principled evaluation}, we deliberately choose \emph{Speech} samples labeled with only one additional child class~(i.e.,~Male speech, Female speech, Child speech, Babbling) to ensure high-quality ground-truth labels for validation. 
%accuracy above 90\%. 
For the other subsets \newadd{domestic sounds and outside or natural audio}, we consider all samples within each category, regardless of label quality, 
% or multiple labels
to validate the generalizability and robustness of \system{} across varying labeling qualities. \newadd{A brief summary of the dataset size is provided in~\tablename~\ref{tab:datasets}}.

\tablename~\ref{tab:datasets} also describes the number of samples used in our experiments. Following our pre-processing pipeline, the number of samples increases, and their lengths vary ($\leq$10~secs), while the original samples are all 10~secs long. The baseline task discovery framework always uses the complete set of original samples. In contrast, \system{} utilizes the complete set of original samples only for \emph{Speech} data. For other subsets, we randomly select 10K samples to \newadd{have a fair comparison and speed up the overall process of discovering new classes}. 

{\bf Cross-dataset validation.} To evaluate downstream audio classifiers, we use FSD50K~\cite{fonseca2022FSD50K}, a benchmark multi-label audio dataset with clip durations ranging from 0.3 to 30 seconds, labeled based on the AudioSet ontology. Note that we \emph{exclusively} use the FSD50K dataset to evaluate the performance of downstream audio classifiers to ensure the validity of \system{} on a completely held-out dataset. 

\subsection{Implementation}
\label{impl}
%To discover new classes from the audio datasets, we extract 64x64~mel-frequency cepstral coefficients~(MFCCs) from the audio samples. 
We used raw audio samples, all 10 seconds in length, for the baseline. For \system{}, we used pre-processed audio samples and applied zero-padding to extend samples shorter than 10 seconds. Using a 25~ms window with a 10~ms overlap, we extracted 64~mel frequency bands and computed mel-frequency cepstral coefficients~(MFCCs), resulting in approximately 1K frames for a 10-second clip. All frames were then resampled to 64~time steps using linear interpolation, ensuring consistent temporal resolution across all samples. %For \system{}, these MFCCs are extracted after the signal pre-processing step~(see \figurename~\ref{fig:taskovery}). \newadd{As the baseline task discovery approach~\cite{atanov2022task} is only designed for images without any pre-processing, we directly use the raw audio data for it.} 
For the baseline approach~\cite{atanov2022task}, we use the official implementation available~\cite{TaskDiscovery}. 
We use default parameters for the task discovery framework and consider the discovered task after the fourth epoch, where the AS typically stabilizes at~0.85 or higher. This threshold is determined by empirical analysis of AS on human-labeled tasks using the baseline task discovery framework. We choose the top~20 samples per class based on the softmax probabilities to annotate the discovered classes without human intervention. These are then fed to a pre-trained audio classifier, YAMNet~\cite{yamnet}. We finally take the top~10 predictions from YAMNet as \newadd{the newly discovered classes}.

We use a Random Forest model with 10 estimators for the downstream audio classifier. Original audio samples are used as input features for cross-validation on AudioSet, while 10-second windows are used for testing on the held-out FSD50K dataset.  
% since FSD50K data ranges from 0.3 to 30 seconds, and we use 10-second samples. 
Each evaluation uses 100 samples per class.

\subsection{Baseline Comparison}
\label{eval:baselinecomparison}
%\ts{R7: It is unclear what exactly the difference between the proposed and baseline algorithms are. Is it only the audio front-end, or also the use of the clarity metric? In the latter case, a proper ablation study could help the reader understand which factor contributes the most.}
\begin{figure*}
    \centering
    \includegraphics[width=0.8\textwidth, keepaspectratio]{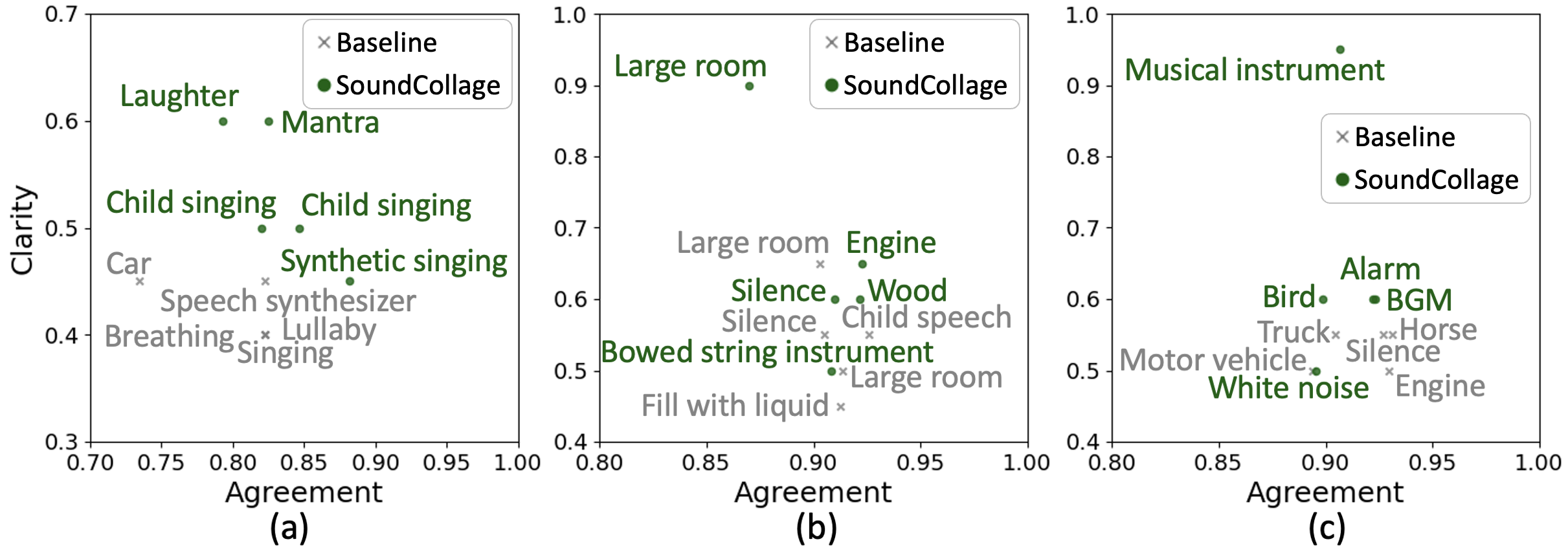}
    % \vspace*{-2.1em}
    \caption{\textbf{Top-5 discovered classes ranked according to clarity for -- (a) speech, (b) domestic, and (c) natural sounds subsets from AudioSet}. Our approach discovers new classes with both higher clarity and agreement.}
  \label{fig:main_perf}
  \vspace{-0.5cm}
\end{figure*}
\begin{table}[t]
\caption{\textbf{\system{} vs. Baseline Task-discovery~\cite{atanov2022task}.}}
\label{tab:clarity}
\resizebox{\columnwidth}{!}{
\scriptsize
\begin{tabular}{cccc}
\Xhline{2\arrayrulewidth}
\textbf{Dataset} & \textbf{System} & \textbf{Clarity} & \textbf{AS} \\ \hline\hline
\multirow{3}{*}{Speech} & Baseline & 0.302 ± 0.070 & 0.803 ± 0.067 \\ \cline{2-4} 
 & Ours Comp\#1 & \textbf{0.306 ± 0.116} & \textbf{0.804 ± 0.095} \\ \cline{2-4} 
 & Ours Comp\#2 & 0.286 ± 0.099 & 0.710 ± 0.104 \\ \hline
\multirow{3}{*}{\begin{tabular}[c]{@{}c@{}}Domestic \\ sounds, \\ home sounds\end{tabular}} & Baseline & 0.336 ± 0.108 & 0.898 ± 0.033 \\ \cline{2-4} 
 & Ours Comp\#1 & 0.350 ± 0.075 & \textbf{0.914 ± 0.026} \\ \cline{2-4} 
 & Ours Comp\#2 & \textbf{0.367 ± 0.160} & 0.903 ± 0.028 \\ \hline
\multirow{3}{*}{\begin{tabular}[c]{@{}c@{}}Outside \\ rural or \\ natural\end{tabular}} & Baseline & 0.342 ± 0.115 & \textbf{0.915 ± 0.024} \\ \cline{2-4} 
 & Ours Comp\#1 & \textbf{0.372 ± 0.150} & 0.914 ± 0.018 \\ \cline{2-4} 
 & Ours Comp\#2 & 0.355 ± 0.086 & 0.903 ± 0.023 \\ \Xhline{2\arrayrulewidth}
\end{tabular}
}
\end{table}

\begin{table}[t]
\caption{\textbf{Diversity of the discovered classes}. \newadd{One of the most discovered class for the baseline approach was \textit{Silence}, whereas our approach discovers more unique classes.}}
\label{tbl:diversity}
\centering
\resizebox{.48\textwidth}{!}{
\begin{tabular}{cccc}
\Xhline{2\arrayrulewidth}
\textbf{Dataset} & \textbf{System} & \multicolumn{1}{c}{\textbf{\begin{tabular}[c]{@{}c@{}}Proportion of\\ \textit{Silence} Class (\%)\end{tabular}}} & \multicolumn{1}{c}{\textbf{\begin{tabular}[c]{@{}c@{}}Total \# of \\ Unique Classes\end{tabular}}} \\ \hline\hline
\multirow{2}{*}{Speech (single label)} & Baseline & 6.3 & 24 \\ \cline{2-4} 
 & Ours & \textbf{0.0} & \textbf{78} \\ \hline
\multirow{2}{*}{\begin{tabular}[c]{@{}c@{}}Domestic sounds,\\ home sounds\end{tabular}} & Baseline & 9.4 & 27 \\ \cline{2-4} 
 & Ours & \textbf{1.6} & \textbf{52} \\ \hline
\multirow{2}{*}{\begin{tabular}[c]{@{}c@{}}Outside rural \\ or natural\end{tabular}} & Baseline & 9.4 & 25 \\ \cline{2-4} 
 & Ours & \textbf{0.0} & \textbf{53} \\ \Xhline{2\arrayrulewidth}
\end{tabular}
}
\end{table}
% We compare \system{}'s performance to the baseline task discovery approach~\cite{atanov2022task}. 
We use clarity (see Eqn.~\ref{eqn:clarity}) and AS (see Eqn.~\ref{eqn:as}) as the primary measures for evaluating class discovery across all the discovered tasks and report the average in \tablename~\ref{tab:clarity}. Our approach generally outperforms the baseline in clarity, demonstrating an average improvement of $0.013$, while remaining in close competition with the baseline in terms of AS. However, in the speech Comp\#2, \system{} underperforms because we selectively used samples originally annotated with a single label, as mentioned in Section~\ref{sec:datasets}, resulting in fewer non-vocal background audio signals. Nevertheless, \textbf{our method achieves higher clarity scores even with fewer original samples} from each class. This is particularly evident for the `outside rural or natural sounds,' where the baseline had \newadd{the entirety of the} 12K samples compared to our \newadd{selected} 10K samples.

We further investigate the classes and labels discovered using the pre-trained audio event model, YAMNet. \figurename~\ref{fig:main_perf} illustrates the top~5 clarity tasks with associated labels. Tasks discovered by \system{} generally exhibit superior clarity and AS.  
\textbf{Our approach also discovers a more diverse set of classes} from the audio data while the baseline redundantly discovers \emph{Silence} (see \tablename~\ref{tbl:diversity}). %Moreover, the baseline redundantly discovers \textit{Silence}. 
This \newadd{quantitative} analysis shows that our pre-processing pipeline extracts semantically meaningful audio components, leading to class boundaries with higher clarity. 
% as measured by clarity, with comparable class generalizability, as indicated by AS.}

We conduct an additional quantitative analysis by comparing the accuracy of downstream audio classifiers (Section~\ref{impl}) trained on samples labeled with the new classes. 
% Specifically, we perform two experiments. 
We assess the classifier performance with 5-fold cross-validation using samples of the highest clarity class across all the discovered tasks. \textbf{The analysis shows that \system{} offers higher classifier performance with consistent improvements of up to 34.7\%} (see \tablename~\ref{tab:downstream_cross}). This is because (1)~the baseline approach lacks a signal pre-processing method tailored for audio data, 
% to separate sound sources, 
and (2)~its discovered classes have lower clarity despite having higher AS.

\subsection{\newadd{Cross-data Quantitative Analysis}}
\label{eval:qual}
Motivated by the aforementioned observation, we further investigate the performance of the downstream classifier on a held-out test set from the FSD50K dataset~\cite{fonseca2022FSD50K} for 
\emph{the common} classes with clarity $\geq0.50$. The results described in~\tablename~\ref{tab:downstream} demonstrate a similar trend, wherein the classifier trained on labeled samples generated by \system{} outperforms the classifier trained on labeled samples provided by the baseline, \hredit{with accuracy improvements up to $4.5\%$.} \textbf{This quantitative analysis shows that the output dataset labeled with new classes by \system{} enables training downstream classifiers with higher accuracy than the baseline.}
\begin{table}[t]
\caption{\textbf{Downstream classification performance on AudioSet.} Each class is the \emph{highest clarity} class among all classes discovered from the original dataset. Labels indicate the semantic meaning assigned by YAMNet.}
\label{tab:downstream_cross}
\resizebox{\columnwidth}{!}{
\begin{tabular}{ccccccc}
\Xhline{2\arrayrulewidth}
\textbf{Original} & \textbf{System} & \textbf{Acc (\%)} & \textbf{Prec (\%)} & \textbf{Rec (\%)} & \textbf{F1 (\%)} & \textbf{Label} \\ \hline\hline
\multirow{2}{*}{Speech} & Baseline & 51.8 ± 4.6 & 51.7 ± 4.8 & 51.8 ± 4.6 & 51.4 ± 5.0 & \begin{tabular}[c]{@{}c@{}}Speech\\ synthesizer\end{tabular} \\ \cline{2-7} 
 & \textbf{Ours} & \textbf{75.6 ± 13.0} & \textbf{77.7 ± 14.1} & \textbf{75.6 ± 13.0} & \textbf{74.0 ± 13.8} & Mantra \\ \hline
\multirow{2}{*}{\begin{tabular}[c]{@{}c@{}}Domestic\\ sounds,\\ home sounds\end{tabular}} & Baseline & 56.2 ± 10.8 & 53.7 ± 16.0 & 56.2 ± 10.8 & 54.1 ± 13.4 & Laughter \\ \cline{2-7} 
 & \textbf{Ours} & \textbf{90.9 ± 5.3} & \textbf{91.1 ± 5.2} & \textbf{90.9 ± 5.3} & \textbf{90.9 ± 5.3} & \begin{tabular}[c]{@{}c@{}}Inside, large\\ room or hall\end{tabular} \\ \hline
\multirow{2}{*}{\begin{tabular}[c]{@{}c@{}}Outside\\ rural\\ or natural\end{tabular}} & Baseline & 62.2 ± 19.7 & 63.1 ± 21.4 & 62.2 ± 19.7 & 59.3 ± 21.1 & Silence \\ \cline{2-7} 
 & \textbf{Ours} & \textbf{91.6 ± 4.2} & \textbf{92.1 ± 4.8} & \textbf{91.6 ± 4.2} & \textbf{91.2 ± 4.6} & \begin{tabular}[c]{@{}c@{}}Musical\\ instrument\end{tabular} \\  \Xhline{2\arrayrulewidth}
\end{tabular}
}
\end{table}

% \begin{table}[]
% \begin{tabular}{ccccccc}
% \hline
% \textbf{Original} & \textbf{System} & \textbf{Discovered} & \textbf{Acc (\%)} & \textbf{Prec  (\%)} & \textbf{Rec  (\%)} & \textbf{F1 (\%)} \\ \hline\hline
% \multirow{2}{*}{Speech} & Baseline & \begin{tabular}[c]{@{}c@{}}Speech\\ synthesizer\end{tabular} & 51.8 ± 4.6 & 51.7 ± 4.8 & 51.8 ± 4.6 & 51.4 ± 5.0 \\ \cline{2-7} 
%  & Ours & Mantra & \textbf{75.6 ± 13.0} & \textbf{77.7 ± 14.1} & \textbf{75.6 ± 13.0} & \textbf{74.0 ± 13.8} \\ \hline
% \multirow{2}{*}{\begin{tabular}[c]{@{}c@{}}Domestic \\ sounds \\ home sounds\end{tabular}} & Baseline & Laughter & 56.2 ± 10.8 & 53.7 ± 16.0 & 56.2 ± 10.8 & 54.1 ± 13.4 \\ \cline{2-7} 
%  & Ours & \begin{tabular}[c]{@{}c@{}}Inside, large \\ room or hall\end{tabular} & \textbf{90.9 ± 5.3} & \textbf{91.1 ± 5.2} & \textbf{90.9 ± 5.3} & \textbf{90.9 ± 5.3} \\ \hline
% \multirow{2}{*}{\begin{tabular}[c]{@{}c@{}}Outside \\ rural \\ or natural\end{tabular}} & Baseline & Silence & 62.2 ± 19.7 & 63.1 ± 21.4 & 62.2 ± 19.7 & 59.3 ± 21.1 \\ \cline{2-7} 
%  & Ours & \begin{tabular}[c]{@{}c@{}}Musical \\ instrument\end{tabular} & \textbf{91.6 ± 4.2} & \textbf{92.1 ± 4.8} & \textbf{91.6 ± 4.2} & \textbf{91.2 ± 4.6} \\
% \hline
% \end{tabular}
% \end{table}

\begin{table}[t]
\caption{\textbf{Downstream classification performance on held-out data from FSD50K.} We report classification results only for \emph{common} classes discovered by both the baseline and \system{}, present in FSD50K with a clarity $\geq0.50$.
}
\label{tab:downstream}
\resizebox{\columnwidth}{!}{
\scriptsize
\begin{tabular}{ccccccc}
\Xhline{2\arrayrulewidth}
\multicolumn{1}{c}{\textbf{Label}} &  \multicolumn{1}{c}{\textbf{System}} & \multicolumn{1}{c}{\textbf{Acc}} & \multicolumn{1}{c}{\textbf{Prec}} & \multicolumn{1}{c}{\textbf{Rec}} & \multicolumn{1}{c}{\textbf{F1}} & \multicolumn{1}{c}{\textbf{Clarity}} \\ \hline\hline
\multirow{2}{*}{\begin{tabular}[c]{@{}c@{}}Motor\\ vehicle\end{tabular}} & Baseline & 0.570 & \textbf{0.592} & 0.450 & 0.511 & \textbf{0.50} \\ \cline{2-7} 
 & Ours & \textbf{0.605} & 0.567 & \textbf{0.890} & \textbf{0.693} & \textbf{0.50} \\ \hline
\multirow{2}{*}{Engine} & Baseline & 0.595 & \textbf{0.732} & 0.300 & 0.426 & 0.50 \\ \cline{2-7} 
 & Ours & \textbf{0.640} & 0.604 & \textbf{0.810} & \textbf{0.692} & \textbf{0.65} \\\Xhline{2\arrayrulewidth}
\end{tabular}
}
\end{table}

\subsection{\newadd{Clarity vs Agreement Score}}
We observe no statistically significant correlation (p-value $>0.2$) between the clarity and the AS for \system{} as well as the baseline. This is because clarity measures how semantically well-defined the discovered task (or the class boundary) is, whereas AS as defined~\cite{atanov2022task} is a similarity between two neural networks. However, as a high AS depicts that the discovered class boundaries are highly generalizable, we observe that the tasks exhibiting a high clarity also exhibit high agreement scores, but the converse is not true. This shows that \textbf{clarity is a more judiciously designed measure for defining the semantic clarity of the discovered tasks than the AS.}

% {\bf Impact of Dataset Size.}
% \begin{figure}[t]
%     \centering
%     \includegraphics[width=0.85\columnwidth,keepaspectratio]{figures/num_data_agreement_clarity.png}
%     \caption{\textbf{Impact of dataset size on the clarity and agreement scores of the discovered tasks.}}
%     \label{fig:dataset_size_agreement_rel}
% \end{figure}
% We investigate the impact of dataset size by varying the total number of samples obtained after the signal processing pipeline that is eventually provided as input to the task discovery sub-module for our framework. From \figurename~\ref{fig:dataset_size_agreement_rel}, we observe
% increasing trends in clarity and agreement score as the dataset size grows in the speech subset of the AudioSet \ds{agreement goes up with more data but clarity not really..}. Notably, \system{} achieves a comparable performance with a much smaller sample size, even with just 20\% of the samples. At the same time, the baseline task discovery has empirically shown a dependency on sample size in AS~\cite{atanov2022task} \ds{we don't see that on the figure. On 100\% both methods look identical. Mention why we cannot run the same experiment for TD}.

% \mo{It is very important to explain the limitations of our work before concluding it. This always prevent a reviewer to come back and mention a limitation as a reason to reject the paper.}

\subsection{Limitations and Future Work.} 
One concern is related to the limitations of the pre-trained audio event classification model in finding all potential meaningful classes. We aim to choose audio-foundation models as an emerging alternative~\cite{chung2021w2vbert, yang2023uniaudio}. Additionally, the end-to-end process of discovering new classes consumes significant time and resources. Thus, another research direction would be redesigning the task discovery algorithm to make it more efficient.

% to address the limitations which still exist when relying on pre-trained audio event classification models for identifying new classes, another viable alternative is to \newadd{use} audio foundation models~\cite{chung2021w2vbert, yang2023uniaudio} to broaden the understanding of hidden classes. 

\section{Conclusion}
We show that current unsupervised task discovery algorithms must be adapted specifically to address the nuances present in the audio data. To this end, we explore various signal pre-processing techniques and potential automated annotation methods. Our findings are integrated into a unified framework, \system{}, designed to facilitate the automated discovery of new classes in audio datasets. Experimental results show that \system{} identifies more unique classes, generally providing improved clarity and consistency compared to the baseline. Additionally, our framework boosts the performance of downstream classifiers. We envision \system{} as a tool to enhance dataset reusability by augmenting samples and expanding their label spaces.

\clearpage
\bibliographystyle{IEEEbib}
\bibliography{taskdiscovery}

% \clearpage
% \begin{NoHyper}
% \input{sections/author_response}
% \end{NoHyper}

\end{document}